\documentclass[12pt,twoside]{article}
\usepackage{amsmath}
\usepackage{amssymb}
\usepackage{array}
\usepackage{graphicx}
\usepackage{geometry}
\usepackage{graphics}
\usepackage{enumerate}
\usepackage[utf8]{inputenc}

\begin{document}
\vskip7cm\noindent
\setcounter{page}{1}
\font\tinyfont=cmr8
\font\headd=cmr8
\pagestyle{myheadings}
\font\bigfont=cmssbx10 at 18pt
\font\tinyfont=cmr8
\font\bigfont=cmssbx10 at 18pt
\begin{center}
{\bf A Note on Mathai's Entropy Measure}
\end{center}

\vskip.1cm\begin{center}Hans J. Haubold\\
Office for Outer Space Affairs, United Nations,\\
 Vienna International Centre, Vienna, Austria\\
hans.haubold@gmail.com\\
\end{center}

\vskip.1cm\noindent{\bf Abstract}

\vskip.1cm In a paper [8] the authors classify entropy into three categories, as a thermodynamics quantity, as a measure of information production and as a means of statistical inference. An entropy measure introduced by Mathai falls into the second and third categories. It is shown that this entropy measure is the same whether the variables involved are real or complex scalar, vector or matrix variables. If the entropy measure is optimized under some moment-like conditions then one can obtain various types of densities which are applicable in different areas. Unlike Tsallis' entropy [9], it does not need an intermediary escort distribution to yield the desired results. Calculus of variation can be directly applied to obtain the desired results under Mathai's entropy. Tsallis' entropy, which is the basis of the area of non-extensive statistical mechanics, is a modified version of the $\alpha$-generalized entropy of Havrda-Charvat considered in [7]. Various types of distributions that can be obtained through optimization of Mathai's entropy,  are illustrated in this paper.

\vskip.3cm\noindent{\bf Subject classification MSC2020:} 94A17, 62810, 62E15, 62H10

\vskip.3cm\noindent{\bf 1.\hskip.3cm Introduction}
\vskip.3cm
In this paper, small letters $x,y,$ etc will be used to denote real scalar variables whether mathematical variables or random variables. Capital letters $X,Y,$ etc will be used to denote vector/matrix variables whether square or rectangular matrices are involved. Constant scalars will be denoted by $a,b,$ etc and vector/matrix constants will be denoted by $A,B,$ etc. Variables in the complex domain will be denoted with a tilde such as $\tilde{x},\tilde{y},\tilde{X},\tilde{Y},$ etc. No tilde will be used on scalar or matrix constants in the complex domain. Let
$$M_{\alpha}(f)=\frac{\int_X[f(X)]^{1+\frac{q-\alpha}{\eta}}{\rm d}X-1}{\alpha-q},\alpha\ne q\eqno(1.1)
$$where $f(X)$ is a real-valued scalar function of $X$ and $X$ may be scalar, vector or matrix in the real or complex domain, such that $f(X)\ge 0$ for all $X$ and $\int_Xf(X){\rm d}X=1$, where ${\rm d}X$ stands for the wedge product of the distinct differentials involved in $X$. For example, if $X$ is a $p\times 1$ real vector with $X'=[x_1,...,x_p]$ where the $x_j$'s are real scalar variables and the prime denotes the transpose, then ${\rm d}X={\rm d}X'={\rm d}x_1\wedge...\wedge{\rm d}x_p$. If $X=(x_{ij})=X'$ (symmetric) is a $p\times p$ real matrix then there are only $p(p+1)/2$ distinct elements and then ${\rm d}X=\wedge_{i\ge j}{\rm d}x_{ij}$. Since $1=\int_Xf(X){\rm d}X$ we may rewrite (1.1) as the following:
$$M_{\alpha}(f)=\int_Xf(X)\left\{\frac{[f(X)]^{\frac{q-\alpha}{\eta}}-1}{\alpha-q}\right\}{\rm d}X=E\left\{\frac{[f(X)]^{\frac{q-\alpha}{\eta}}-1}{\alpha-q}\right\}
\eqno(1.2)
$$where $\alpha\ne q,\eta>0,q$ is real scalar, $\alpha$ is a parameter, and $E(\cdot)$ denotes the expected value of $(\cdot)$. In (1.2), $q$ is an anchoring point, including zero, the departure of $\alpha$ from this fixed point $q$ is measured in $\eta$ units. An earlier version of (1.2) with $q=1,\eta=1$ was introduced by Mathai in 2005 soon after the publication of the pathway model [2] in order to avoid the use of escort density in the use of Tsallis' entropy, and reported in [3]. Then by 2009 $M_{\alpha}(f)$ was modified to bring it in the structure of (1.2), which was reported in [4]. Observe that
$$\lim_{\alpha\to q}\frac{[f(X)]^{\frac{q-\alpha}{\eta}}-1}{\alpha-q}=\lim_{\alpha\to q}\frac{\frac{\partial}{\partial \alpha}\{[f(X)]^{\frac{q-\alpha}{\eta}}-1\}}{\frac{\partial}{\partial\alpha}[\alpha-q]}=-\frac{1}{\eta}\ln f.\eqno(1.3)
$$Thus, in the limit when $\alpha\to q$, $M_{\alpha}(f)\to S_{\alpha}(f)$ where $S_{\alpha}(f)=K\int_xf(x)\ln f(x){\rm d}x$ is Shannon's entropy where $x$ is real scalar and $K$ is a constant, whereas (1.3) covers all types of variables, scalar, vector, matrix, real or complex. We consider only the continuous version or densities only. For the discrete case of the multinomial probabilities $p_i>0,i=1,...,k,p_1+...+p_k=1$ the entropy measure corresponding to $M_{\alpha}(f)$ of (1.1) is the following, again denoted by $M_{\alpha}$:
$$M_{\alpha}=\frac{\sum_{j=1}^kp_j^{1+\frac{q-\alpha}{\eta}}-1}{\alpha-q},\alpha\ne q.\eqno(1.4)
$$When taking the limit, if any $p_i$ is assumed to be zero or if an impossible event is included then $0\ln 0$ is assumed to be zero.

\vskip.3cm\noindent{\bf 2.\hskip.3cm Optimization of $M_{\alpha}(f)$ in the Scalar Case}
\vskip.3cm
Let us consider a real scalar positive variable $x>0$ and a real-valued scalar function $f(x)$ of $x$ such that $f(x)\ge 0$ and $\int_{-\infty}^{\infty}f(x){\rm d}x=1$ or $f(x)$ is a statistical density. For the real scalar positive variable case, let us consider the optimization of $M_{\alpha}(f)$ of (1.1) under the following conditions:
\vskip.2cm\noindent (i):~~$E[x^{\gamma_1(q-\alpha)}]=\int_{-\infty}^{\infty}x^{\gamma_1(q-\alpha)}f(x){\rm d}x=\mbox{ a fixed quantity}$ over all functional $f$.
\vskip.1cm\noindent (ii):~~$E[x^{\gamma_1(q-\alpha)+\delta}]=\mbox{ a fixed quantity}$ over all functional $f$ for some $\gamma_1$ and $\delta$.
\vskip.1cm\noindent
Note that when $\alpha=q$, condition (i) is the same as $\int_xf(x){\rm d}x=1$. When $\alpha=q$ and $\delta=1$ the second condition says that the first moment or $E[x]$ is fixed. This can be interpreted as the law of conservation of energy when $f(x)$ is an energy density. The conditions imposed are moment-like and hence physical interpretations can be given in terms of moments. If Calculus of variation techniques are used for the optimization then the Euler function and Euler equation are the following, where $\lambda_1$ and $\lambda_2$ are Lagrangian multipliers:
\begin{align*}
g(f)&=f^{1+\frac{q-\alpha}{\eta}}-\lambda_1 x^{\gamma_1(q-\alpha)}f+\lambda_2x^{\gamma_1(q-\alpha)+\delta}f,x>0\\
\frac{\partial}{\partial f}g(f)=0&\Rightarrow (1+\frac{q-\alpha}{\eta})f^{\frac{q-\alpha}{\eta}}-\lambda_1x^{\gamma_1(q-\alpha)}+\lambda_2x^{\gamma_1(q-\alpha)+\delta}\\
&\Rightarrow f=\lambda x^{\gamma_1\eta}[1-a(q-\alpha)x^{\delta}]^{\frac{\eta}{q-\alpha}}\tag{2.1}\end{align*}where $\lambda$ is the normalizing constant, $\frac{\lambda_2}{\lambda_1}$ is taken as $a(q-\alpha)$ for some $a>0$ for convenience. Then for $a>0,\eta>0,\alpha<q,\gamma=\gamma_1\eta$ and $1-a(q-\alpha)x^{\delta}>0$,
$$f_1(x)=c_1x^{\gamma}[1-a(q-\alpha)x^{\delta}]^{\frac{\eta}{q-\alpha}},\alpha<q\eqno(2.2)
$$is a generalized type-1 beta density. Then for $\alpha>q$ this (2.2) switches into the density
$$f_2(x)=c_2x^{\gamma}[1+a(\alpha-q)x^{\delta}]^{-\frac{\eta}{\alpha-q}},\alpha>q,x>0\eqno(2.3)
$$which is a generalized type-2 beta density. When $\alpha\to q$ then both (2.2) and (2.3) go to
$$f_3(x)=c_3x^{\gamma}{\rm e}^{-a\eta x^{\delta}},a>0.\eta>0,x>0
$$which is a generalized gamma density, where $c_1,c_2,c_3$ are the corresponding normalizing constants. Thus, the three densities $f_j(x),j=1,2,3$ are connected through the pathway parameter $\alpha$ and the model in (2.2) or (2.3) is called a pathway model for the real scalar positive variable case. Replace $x$ by $|x|$ in $f_1(x),f_2(x),f_3(x)$ to cover the whole real line, $-\infty<x<\infty$. In the type-1 case the support of the density is finite. When $\tilde{x}$ is in the complex domain replace $x$ in $f_j(x)$ by $\tilde{x}^{*}\tilde{x}$ where $\tilde{x}^{*}$ is the conjugate of $\tilde{x}$. Then (2.4) gives an extension of Maxwell-Boltzmann and Rayleigh densities to the scalar complex variable case and available from (1.1) by optimization. This extension to the complex domain may not be available in the literature. Observe that $f_j(x),j=1,2,3$ cover a wide range of densities in standard use in Statistics, Physics, Engineering and other areas, all coming from the optimization of (1.1).

\vskip.3cm\noindent{\bf 3.\hskip.3cm Optimization in the Real and Complex Multivariate Case}
\vskip.3cm
Multivariate case usually means a collection of scalar variables. Let $X$ be a $p\times 1$ vector with the real elements $x_1,...,x_p$. Then a norm of $X$, denoted by $\Vert X\Vert$ can be defined in various ways which will also be distance of $X$ from the origin $O$. Let us consider the Euclidean norm. Then $\Vert X\Vert^2=x_1^2+...+x_p^2=X'X$ when $x_j$'s are real. In statistical problems one may wish to get rid off the effects of joint variations among the components. Then we consider a generalized norm. The generalized Euclidean norm is $\Vert X\Vert^2=X'\Sigma^{-1}X$ where $\Sigma>O$ is the covariance matrix in $X$. This positive definite quadratic form also has statistical interpretations. $X'\Sigma^{-1}X=c>0$ is called the ellipsoid of concentration, the probability content in $X'\Sigma^{-1}X\le c$ is the probability around the origin $X=O$ for the distribution of $X$. For convenience we may consider densities which are functions of $X'AX,A>O$ which will cover the square of the ordinary distance from the origin as well as the square of the generalized distance for $A=I$ and $A=\Sigma^{-1}$ respectively. Then the conditions will be in terms of the moments of $X'AX$. Consider the following conditions:
\vskip.2cm\noindent (iii):~~$E[X'AX]^{\gamma_1(q-\alpha)}=\mbox{ fixed}$ over all functional $f$, and
\vskip.1cm\noindent (iv)~~$E[X'AX]^{\gamma_1(q-\alpha)+\delta}=\mbox{ fixed}$ over all functional $f$.
\vskip.1cm\noindent
Let us optimize the entropy in (1.1) where $X$ is now a real $p\times 1$ vector. Then proceeding as in the real scalar case we end up with the following density:
$$
f_4(X)=c(X'AX)^{\gamma_1\eta}[1-bX'AX]^{\frac{\eta}{q-\alpha}}.$$Take $\gamma_1\eta=\gamma, b=a(\gamma-\alpha),a>0.$ Then for $\alpha<q,\alpha>q,\alpha\to q$ we have the following three pathway densities.

\begin{align*}
f_5(X)&=C_1(X'AX)^{\gamma}[1-a(q-\alpha)(X'AX)^{\delta}]^{\frac{\eta}{q-\alpha}},\alpha<q\tag{3.1}\\
f_6(X)&=C_2(X'AX)^{\gamma}[1+a(\alpha-q)(X'AX)^{\delta}]^{-\frac{\eta}{\alpha-q}},\alpha>q\tag{3.2}\\
f_7(X)&=C_3(X'AX)^{\gamma}{\rm e}^{-a\eta (X'AX)^{\delta}},a>0.\eta>0\tag{3.3}\end{align*}Then $f_5,f_6,f_7$ are the pathway densities in the real vector variable case, where $C_1,C_2,C_3$ are the corresponding normalizing constants. Note that since $A>O$ we can put $X'AX$ as a sum of squares $y_1^2+...+y_p^2$. Then $f_5$ or $f_6$, from where one can go to $f_j,j=5,6,7$, gives the multivariate version of the pathway density. For $\gamma=0$ such forms are the generalized type-1 beta, type-2 beta and gamma distributed isotropic random points in geometrical probability problems, see [1]. Such models are also multivariate analogues of popular densities used in the area of reliability analysis, see also [5]. Observe that (3.3) also gives the real multivariate generalization of Maxwell-Boltzmann and Rayleigh densities, see[6]. When $(X'AX)=Y'Y$ for a $p\times 1$ vector $Y$, then (3.3) also provides the isotropic forms of the Maxwell-Boltzmann and Rayleigh densities which are invariant under orthonormal transformations or rotations of the axes of coordinates. (3.1)-(3.3) also give densities of quadratic forms under pathway model. Observe that here $-\infty<x_j<\infty,j=1,...,p$. Also note that (3.1)-(3.3) belong to the family of  elliptically contoured distributions and the corresponding forms in $Y$ belong to the spherically symmetric family of distributions.
\vskip.2cm
We can extend the densities in (3.1)-(3.3) to the complex domain also. Replace $X'$ by $\tilde{X}^{*}$ in (3.1)-(3.3) to obtain the corresponding versions for the complex case, where $\tilde{X}^{*}$ means the complex conjugate transpose of $\tilde{X}$. Then in the canonical case $$\tilde{Y}^{*}\tilde{Y}=|\tilde{y}_1|^2+...+|\tilde{y}_p|^2,\tilde{y}_j=t_{j1}+it_{j2},|\tilde{t}_j|^2=t_{j1}^2+t_{j2}^2
$$where $t_{j1}$ and $t_{j2}$ are real quantities and $i=\sqrt{(-1)}$. Evaluation of the normalizing constants as well as other integrals connected with the complex versions of (3.1)-(3.3) can be done by using a $2p$-dimensional polar coordinate transformation. We can also use other results from Special Functions. We have ${\rm d}S$ available in terms of ${\rm d}X$ where $S=\tilde{Y}\tilde{Y}^{*}$ and this result can be used to obtain such results very quickly. Complex version of (3.3) also provides generalized complex multivariate cases of Maxwell-Boltzmann and Rayleigh densities. Basic forms are available for $\delta=1$. None of these forms, even for the complex scalar case $p=1$, may be available in the literature yet.

\vskip.3cm\noindent{\bf 4.\hskip.3cm Matrix-variate Generalizations in the Real and Complex Cases}
\vskip.3cm
To start with consider a real $p\times p$ positive definite matrix $X>O$ in (1.1).  Consider the optimization of (1.1) under the following conditions, where $|X|$ denotes the determinant of the real $p\times p$ matrix $X$:
\vskip.2cm\noindent (v):~~$E[|X|^{\gamma_1(q-\alpha)}|I-bX|^{\gamma_2}=\mbox{ fixed}$ over all functional $f$.\\
Here, when $\gamma_2=0$ the condition is on $|X|$ and when $\gamma_1=0$ the condition is on $|I-bX|$ where $b$ is a scalar constant. Then, optimizing (1.1) under the restriction in (v) and following through the steps in earlier sections we have the density of the following form:
$$f_8(X)=c|X|^{\gamma_1\eta}|I-bX|^{\frac{\gamma_1\eta}{q-\alpha}}$$where $c$ is a constant. Take $b=a(q-\alpha),a>0,\gamma_1\eta=\gamma, \gamma_2\eta=\eta_1$ for $\alpha<q$ then $f_8$ changes to the following three densities for the cases $\alpha<q,\alpha>q,\alpha\to q$:
\begin{align*}
f_9(X)&=C_9|X|^{\gamma}|I-a(q-\alpha)X|^{\frac{\eta_1}{q-\alpha}},\alpha<q\tag{4.1}\\
f_{10}(X)&=C_{10}|X|^{\gamma}|I+a(\alpha-q)X|^{-\frac{\eta_1}{\alpha-q}},\alpha>q\tag{4.2}\\
f_{11}(X)&=C_{11}|X|^{\gamma}{\rm e}^{-a\eta_1{\rm tr}(X)}\tag{4.3},\end{align*}
where $C_j,j=9,10,11$ are normalizing constants, $a>0,\eta_1>0,\gamma>-1$ and in (4.2), $I-a(q-\alpha)X>O$ (positive definite) in order to make it a density. Then $f_9$ is the real $p\times p$ matrix-variate generalized type-1 beta density with the parameters ($\gamma+\frac{p+1}{2},\frac{\eta_1}{q-\alpha}+\frac{p+1}{2}$), $f_{10}$ is a generalized $p\times p$ real matrix-variate type-2 beta density with the same parameters as above and $f_{11}$ is a real matrix-variate gamma density. We can show that when $\alpha\to q$ we have $|I-a(q-\alpha)X|^{\frac{\eta_1}{q-\alpha}}$ as well as $|I+a(\alpha-q)X|^{-\frac{\eta_1}{ \alpha-q}}$ going to  ${\rm e}^{-a\eta_1{\rm tr}(X)}$ where ${\rm tr}(\cdot)$ denotes the trace of $(\cdot)$. Note that all the three densities in (4.1)-(4.3) are available from (4.1) or (4.2) via the pathway parameter $\alpha$. Real matrix-variate random points considered in [1] are special cases of the densities in (4.1)-(4.3). The density in (4.3) is  real matrix-variate generalization of the Maxwell-Boltzmann and Rayleigh densities. Such real Maxwell-Boltzmann and Rayleigh densities are considered in [6].
\vskip.2cm
We can have an extension of the results in (4.1)-(4.3) by replacing $X>O$ by $A^{\frac{1}{2}}XA^{\frac{1}{2}}$ where $A>O$ is a constant real positive definite matrix. We can extend the results in (4.1)-(4.3) to the complex domain. In this case replace $|X|$ and $|I-bX|$ by $|{\rm det}(\tilde{X})|$ and $|{\rm det}(I-b\tilde{X})|$ respectively where $|{\rm det}(\cdot)|$ represents the absolute value of the determinant of $(\cdot)$. Absolute value of the determinant of $\tilde{X}$ means $\sqrt{{\rm det}(\tilde{X}^{*}\tilde{X})}$. Then (4.3) extended to the complex domain gives the complex matrix-variate Maxwell-Boltzmann and Rayleigh densities. Such densities may not be available in the literature. The densities in (4.1)-(4.3) can also be extended to the real as well as complex rectangular matrix-variate cases. In this case the $X$ in the entropy (1.1) will be a $p\times m$ rectangular matrix, $p\le m$ and of full rank $p$. Then the condition $(v)$ can be replaced by a condition (vi), where
\vskip.2cm\noindent
(vi):~~$|A^{\frac{1}{2}}XBX'A^{\frac{1}{2}}|^{\gamma_1(q-\alpha)}|I-bA^{\frac{1}{2}}XBX'A^{\frac{1}{2}}|^{\gamma_2}=\mbox{a fixed quantity}$
\vskip.2cm\noindent
over all functional $f$ where $A>O$ is $p\times p$ and $B>O$ is $m\times m$ real positive definite constant matrices. Then in the densities in (4.1)-(4.3) replace $X$ by $A^{\frac{1}{2}}XBX'A^{\frac{1}{2}}$. We can also have an extension to the complex rectangular matrix-variate case. In this case the matrix $X$ in (1.1) is a $p\times m$ rectangular matrix, $p\le m$ and of rank $p$. The condition (vi) will change to (vii), where 
\vskip.2cm\noindent
(vii):~~$|{\rm det}(A^{\frac{1}{2}}\tilde{X}B\tilde{X}^{*}A^{\frac{1}{2}})|^{\gamma_1(q-\alpha)}|{\rm det}(I-bA^{\frac{1}{2}}\tilde{X}B\tilde{X}^{*}A^{\frac{1}{2}})|^{\gamma_2}=\mbox{fixed}$
\vskip.2cm\noindent
over all functional $f$. Then the changes will be to replace $X$ in (4.1)-(4.3) by $A^{\frac{1}{2}}\tilde{X}B\tilde{X}^{*}A^{\frac{1}{2}}$. The techniques of tackling such distributions are illustrated in [6] for the real case. Maxwell-Boltzmann and Rayleigh densities will be available from the complex matrix-variate extended version of (4.3). Such extended results are not available in the literature.

\vskip.3cm\begin{center}{\bf References}\end{center}

\vskip.2cm\noindent [1]~~A.M. Mathai (1999): {\it An Introduction to Geometrical Probability: Distributional Aspects with Applications}, Gordon and Breach, Amsterdam.
\vskip.2cm\noindent [2]~~A.M. Mathai (2005): A pathway to matrix-variate gamma and normal densities, {\it Linear Algebra and its Applications}, {\bf 396}, 317-328.
\vskip.2cm\noindent [3]~~A.M. Mathai and H.J. Haubold (2007): Pathway model, superstatistics, Tsallis statistics and a generalized measure of entropy, {\it Physica A}, {\bf 375}, 110-126.
\vskip.2cm\noindent [4]~~A.M. Mathai and H.J. Haubold (2017):~~ A generalized entropy optimization and Maxwell-Boltzmann densities, {\it European Physical Journal B}, {\bf 91 paper 39}, Doi: org/10.1140/epjb/e2017-80371-5.
\vskip.2cm\noindent [5]~~A.M. Mathai and T. Princy (2017): Analogues of reliability analysis for matrix-variate cases, {\it Linear Algebra and its Applications}, {\bf 532}, 287-311.
\vskip.2cm\noindent [6]~~A.M. Mathai and T. Princy (2017): Multivariate and matrix-variate analogues of Maxwell-Boltzmann and Rayleigh densities, {\it Physica A}, {\bf 468}, 668-676.
\vskip.2cm\noindent [7]~~A.M. Mathai and P.N. Rathie (1975): {\it Basic Concepts in Information Theory and Statistics: Axiomatic Foundations and Applications}, Wiley Halsted, New York.
\vskip.2cm\noindent [8]~~S. Thurner, B. Corominas-Murtea, and R. Hanel (2017): The three faces of entropy for complex systems - information, thermodynamics and the maxent principle, DOI: https://doi.org/10.1103/PhysRevE.96.032124, {\it arXiv:1705-07714v1}.
\vskip.2cm\noindent [9]~~C. Tsallis (1988): Possible generalization of Boltzmann-Gibbs statistics, {\it Journal of Statistical physics}, {\bf 52}, 479-487.

\end{document}